\documentclass{PoS}
\usepackage{amsmath}
\usepackage{epsfig}

\newcommand{\be}{\begin{equation}}
\newcommand{\ee}{\end{equation}}
\newcommand{\bea}{\begin{eqnarray}}
\newcommand{\eea}{\end{eqnarray}}
\newcommand{\nn}{\nonumber}

\def\als{\alpha_{\rm s}}

\def\siml{{\ \lower-1.2pt\vbox{\hbox{\rlap{$<$}\lower6pt\vbox{\hbox{$\sim$}}}}\ }}     
\def\simg{{\
    \lower-1.2pt\vbox{\hbox{\rlap{$>$}\lower6pt\vbox{\hbox{$\sim$}}}}\ }}

\title{Effective field theories for heavy quarkonium at finite temperature}

\ShortTitle{Effective field theories for heavy quarkonium at finite temperature}

\author{\speaker{Antonio Vairo}
\\
        Physik Department, Technische Universit\"at M\"unchen, 85748 Garching, Germany \\
        E-mail: \email{antonio.vairo@ph.tum.de}}

\abstract{We discuss the recent development of effective field theories 
for quarkonium at finite temperature.}

\FullConference{8th Conference Quark Confinement and the Hadron Spectrum \\
		 September 1-6 2008\\
		 Mainz, Germany}

\begin{document}

\section{Introduction}
Experiments in past (SPS), present (RHIC) and future (LHC) colliders 
are attempting to recreate an early condition of the universe known as the quark-gluon plasma, 
where quarks and gluons exist without being bound into hadrons. 
Colliders explore the zero chemical potential region of the QCD phase diagram where  
lattice simulations indicate that a significant increase in the degrees of
freedom happens above a certain critical temperature $T_c \approx 175$ MeV 
(for a recent review see \cite{Satz:2005hx}).

Heavy quarkonium dissociation has been proposed long time ago as a clear probe of the quark-gluon plasma 
formation in colliders through the measurement of the dilepton decay-rate signal \cite{Matsui:1986dk}.
Since higher excited quarkonium states are more weakly bound than lower ones, the expectation is 
that, as the temperature increases, quarkonium will dissociate subsequently from the higher to the lower 
states providing also a dynamical probe of the quark-gluon plasma formation  
(for some recent experimental data see \cite{Leitch08}).

In order to study quarkonium properties in a thermal bath at a temperature $T$, 
the quantity to be determined is the quarkonium potential $V$, which 
dictates, through  the Schr\"odinger equation
\be
E \, \Phi = \left(\frac{p^2}{m}+ V\right)\,\Phi\,,
\label{schroe}
\ee
the real-time evolution of the wave function $\Phi$ of a $Q\bar Q$ pair in the medium.
In the full theory, $V$ must come from a systematic expansion
in $1/m$ (non-relativistic expansion), the leading term being the static potential, and  
in the energy $E$ (ultrasoft expansion). 
The potential  will encode all contributions from scales larger than $E$ and smaller than $m$.
If the temperature lies in this range, the potential will depend on it, if the temperature 
is smaller than or of the same order as $E$, the potential will be temperature independent.

The expansions in $1/m$ and $E$ are best implemented in QCD by means of effective field 
theories (EFTs), very much in the same way as this has been done in order to describe 
quarkonium physics at zero temperature \cite{Brambilla:2004jw}.
In the EFTs, the full dynamics will be more complicated than the Schr\"odinger equation (\ref{schroe}), 
since the EFTs will account both for potential and/or non-potential interactions. 
However, Eq. (\ref{schroe}) will provide the correct leading-order dynamics.

In the last two years, there has been a remarkable progress in constructing EFTs 
for quarkonium at finite temperature and in rigorously defining the quarkonium potential.
In \cite{Laine:2006ns,Laine:2007qy}, the static potential was calculated 
in the regime $T \gg 1/r \simg m_D$, where $m_D$ is the Debye mass and $r$ the quark-antiquark 
distance, by performing an analytical continuation of the Euclidean Wilson loop
to real time. The calculation was done in the weak-coupling resummed perturbation
theory. The imaginary part of the gluon self energy gives an imaginary part to
the static potential and hence a thermal width to
the quark-antiquark bound state. In the same framework, the dilepton
production rate for charmonium and bottomonium was calculated in \cite{Laine:2007gj,Burnier:2007qm}.
In \cite{Beraudo:2007ky}, static particles in real-time formalism were considered 
and the potential for distances $1/r \sim m_D$ was derived for a hot QED plasma. 
The real part of the static potential was found to agree with the
singlet free energy and the damping factor with the one found in \cite{Laine:2006ns}.
In \cite{Escobedo:2008sy}, a study of bound states in a hot QED 
plasma was performed in a non-relativistic EFT framework. 
In particular, the hydrogen atom was studied for temperatures ranging from 
$T\ll m\alpha^2$ to $T\sim m$, where the imaginary part of the
potential becomes larger than the real part and the hydrogen ceases to exist. 
An EFT framework in real time and weak coupling for quarkonium at finite
temperature was developed in \cite{Brambilla:2008cx}; in the rest of the 
presentation, we will follow closely that approach.

\section{Scales and effective field theories}
Quarkonium in a medium is characterized by different energy and momentum scales; 
there are the scales of the non-relativistic bound state ($v$ is the relative heavy-quark velocity):
$m$, the heavy quark mass, $mv$, the scale of the typical inverse 
distance between the heavy quark and antiquark, $mv^2$, the scale of the 
typical binding energy or potential and lower energy scale, and there are the  
thermodynamical scales: the temperature $T$, the inverse of the screening 
length of the chromoelectric interactions, i.e. the Debye mass $m_D$ and lower scales, which 
we will neglect in the following.

\begin{figure}
\makebox[3truecm]{\phantom b}
\put(20,0){\epsfxsize=8truecm \epsfbox{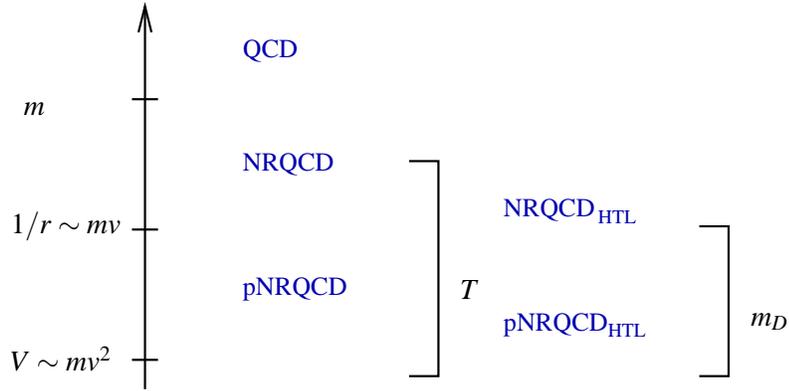}}
\put(-20,105){$m$}
\put(-25,60){$1/r \sim mv$}
\put(-25,8){$V \sim mv^2$}
\put(145,35){$T$}
\put(255,25){$m_D$}
\caption{Quarkonium at finite temperature: energy scales and EFTs.}
\label{figscales}
\end{figure}

If these scales are hierarchically ordered, then we may expand physical observables in the 
ratio of the scales. If we separate explicitly the contributions from the different scales
at the Lagrangian level this amounts to substituting QCD with a hierarchy of EFTs, which are equivalent 
to QCD order by order in the expansion parameters. At zero temperature the EFTs
that follow from QCD by integrating out the scales $m$ and $mv$ are called respectively 
Non-relativistic QCD (NRQCD) and potential NRQCD (pNRQCD), see  \cite{Brambilla:2004jw} for a review.
We assume that the temperature is high enough that $T \gg gT \sim m_D$ holds 
but also that it is low enough for $T \ll m$ and $1/r \sim mv \simg m_D$ to be satisfied, 
because for higher temperature  the bound state ceases to exist.
Under these conditions some possibilities are in order. If $T$ is the next relevant scale after 
$m$, then integrating out $T$ from NRQCD leads to an EFT that we may name NRQCD$_{\rm HTL}$, because 
it contains the hard thermal loop (HTL) Lagrangian \cite{Braaten:1989mz}. Subsequently integrating out 
the scale $mv$ from NRQCD$_{\rm HTL}$ leads to a thermal version of pNRQCD that we may call 
pNRQCD$_{\rm HTL}$. If the next relevant scale after $m$ is $mv$, 
then integrating out $mv$ from NRQCD leads to pNRQCD. If the temperature is larger than $mv^2$, 
then the temperature may be integrated out from pNRQCD leading to a new version of pNRQCD$_{\rm HTL}$.
The hierarchies of scales that lead to these different EFTs are schematically illustrated 
in Fig.~\ref{figscales}. Note that, as long as the temperature is smaller than the scale being 
integrated out, the matching leading to the EFT may be performed putting the temperature to zero.

In the following we will also assume that $v \sim \als$, which is expected to be valid 
for tightly bound states: $\Upsilon(1S)$, $J/\psi$, ...~.

The mass $m$ is the largest scale in the system. This allows to integrate out $m$ 
first and organize the EFTs as expansions in $1/m$. 
The leading order in the $1/m$ expansion corresponds to the static limit of NRQCD: 
\be
{\cal L}  = 
- \frac{1}{4} F^a_{\mu \nu} F^{a\,\mu \nu} 
+ \sum_{i=1}^{n_f}\bar{q}_i\,iD\!\!\!\!\!/\,q_i 
+ \psi^\dagger i D_0 \psi  + \chi^\dagger i D_0 \chi\,,
\label{staticQCD}
\ee
where $\psi$ ($\chi$) is the field that annihilates (creates) the (anti)fermion; 
$q_i$ are $n_f$ light (massless) quark fields. Only longitudinal gluons couple to static quarks.
The relevant scales in static NRQCD are: $1/r$, $V$, ... $T$, $m_D$, ...~.

Since we are interested in the real-time evolution of the heavy quark-antiquark pair,   
it is convenient to modify the contour of the partition function in order to allow for 
real times, see, for instance, \cite{LeBellac:1996}.
In real time, the degrees of freedom double, modifying the propagators into 2 $\times$ 2 matrices.
Despite this, the advantages are that the way in which calculations are carried out is very close to the one 
for $T=0$ EFTs, moreover, in the static quark sector, the second degrees of freedom, labeled ``2'', 
decouple from the physical degrees of freedom, labeled ``1''. The technical reason for this is that 
the $[{\bf S}_{Q}^{(0)}(p)]_{12}$ component of a static quark propagator vanishes, hence  
the unphysical static  quark fields ``2'' never enter in any physical amplitude, i.e. any amplitude 
that has the physical fields ``1'' as initial and final states. 
It is also very convenient to chose the Coulomb gauge: in Coulomb gauge, only transverse 
gluons carry a thermal part, but they do not couple to static quarks. 
Finally, also the static quark-antiquark potential has a 2 $\times$ 2 matrix structure, which reads 
\bea
\left(
\begin{matrix}
&& V
&&0 \\ 
&& -2i \, {\rm Im} \, V
&&\displaystyle - V^*
\end{matrix}
\right) 
.
\eea
In the following, whenever we speak about the potential, we mean the physical one, i.e. 
the entry $V$ in the above matrix.

\section{Static quark antiquark at $T \siml V$}
If the temperature is very low, $T \siml V$, then it does not affect the potential,  
which may be derived by integrating out the scale $1/r$ from (\ref{staticQCD}).
This leads to pNRQCD in the static limit,  whose degrees of freedom are quark-antiquark states 
(color singlet S, color octet O), low energy gluons and light quarks.
The Lagrangian is organized as an expansion in $r$:
\bea
& &\hspace{-12mm}
{\cal L} =  - \frac{1}{4}   F_{\mu\nu}^a F^{\mu\nu\,a} 
+ \sum_{i=1}^{n_f}\bar{q}_i\,iD\!\!\!\!\!/\,q_i 
+ {\rm Tr} \left\{ {\rm S}^\dagger \left( i\partial_0 - V_s \right){\rm S}
+  {\rm O^\dagger} \left( i{D_0} - V_o \right){\rm O}\right\}
\nn \\
& &\hspace{-4mm} 
+ V_A{\rm Tr} 
\left\{  {\rm O^\dagger}{\bf r}\cdot g {\bf E}\,{\rm S}
+ {\rm S^\dagger} {\bf r}\cdot g {\bf E} \,{\rm O} \right\} 
+ \frac{V_B}{2} {\rm Tr} \left\{  {\rm O^\dagger} 
{\bf r} \cdot g {\bf E}\,{\rm O}
+ {\rm O^\dagger} {\rm O } {\bf r} \cdot g{\bf E} \right\} + \cdots\,.
\eea
At leading order in $r$, the singlet decouples from the octet and its equation 
of motion is $(i\partial_0 -$ $V_s){\rm S} =0 $. We may identify $V_s$ and $V_o$ with 
the singlet and octet potentials. They are Coulombic: $\displaystyle V_s(r) = -C_F \frac{\als}{r}$
and $\displaystyle V_o(r) =  \frac{\als}{2N_c\,r}$ at leading order in $\als$ ($N_c =3$, $C_F=4/3$).

\begin{figure}
\centerline{{\epsfxsize=6truecm \epsfbox{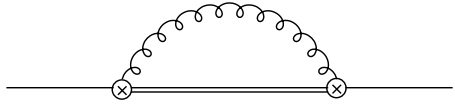}}}
\caption{Single and double lines stand respectively for color singlet and color octet 
quark-antiquark propagators. The curly line stands for the chromoelectric correlator 
$\langle {\bf E}^a(t)\phi_{ab}(t,0){\bf E}^b(0) \rangle$, where $\phi_{ab}$ is a Wilson 
line in the adjoint representation, circles with cross stand for chromoelectric dipole interactions.}
\label{figusT}
\end{figure}

\begin{figure}[ht]
\centerline{{\epsfxsize=4truecm \epsfbox{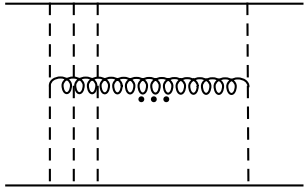}}}
\caption{QCD diagrams responsible for the singlet to octet transition width in a thermal bath.}
\label{figbox}
\end{figure}

Thermal corrections do not affect the potential, but affect the static energy and the decay 
width through loop corrections. The leading correction is carried by the diagram shown in Fig.~\ref{figusT}.
The real part of the diagram gives the following thermal correction to the 
static energy
\be
\delta E = 
\frac{2}{3}\, N_c C_F \, \frac{\als^2}{\pi} \, r \, T^2\, f\left({N_c\als}/{(2rT)} \right)\,,
\ee
where $\displaystyle f(z)$ $=$ $\displaystyle \int_0^\infty dx\, \frac{x^3}{e^x-1}\;$ 
$\displaystyle {\rm P} \frac{1}{x^2-z^2}$ $=$ $\displaystyle \frac{z^2}{2}\left[ \ln \frac{z}{2\pi} 
- {\rm Re} \, \psi\left(i\frac{z}{2\pi}\right) \right]$ $+$ $\displaystyle \frac{\pi^2}{6}$.
The imaginary part of the diagram gives the thermal width 
\be
\Gamma 
= \frac{N_c^3C_F}{6} \, \frac{\als^4}{r}\, n_{\rm B}\left({N_c\als}/{(2r)}\right)\,,
\ee
where $n_{\rm B}(k^0) = 1/(e^{k^0/T}-1)$ is the Bose statistical factor.
Corrections coming from the scale $m_D$ are suppressed by powers of $m_D/T$.
The width $\Gamma$ originates from the fact that thermal fluctuations 
of the medium at short distances may destroy a color-singlet $Q\bar{Q}$ into an octet plus 
gluons. This process is specific of QCD at finite $T$. We will call this
process the singlet to octet break-up phenomenon; in QCD the corresponding diagrams 
are shown in Fig. \ref{figbox}.

In the limiting case $T \ll V$, we have 
\be
\delta E = - \frac{8}{45}\, \pi^3 \, \frac{C_F}{N_c} \, r^3\, T^4 = - \frac{4}{3}\, \pi\, \frac{C_F}{N_c} \, r^3\,  
\langle {\bf E}^a(0)\cdot {\bf  E}^a(0)\rangle_T\,,
\ee
and $\Gamma$ is exponentially suppressed.

\section{Static quark antiquark at $1/r \gg T \gg V$}
In the situation $1/r \gg T \gg V$, integrating out $T$ from pNRQCD modifies pNRQCD 
into a new EFT, pNRQCD$_{\rm HTL}$. With respect to pNRQCD, the Yang--Mills
sector of the pNRQCD$_{\rm HTL}$ Lagrangian gets an additional hard thermal loop part
\cite{Braaten:1989mz}, which modifies, for instance, the longitudinal gluon propagator at $k^0=0$ as
\bea
\frac{i}{{\bf  k}^2}  \to
 \frac{i}{{\bf k}^2 + m_D^2}  
+
\pi \,\frac{T}{|{\bf k}|}\,\frac{m_D^2}{\left({\bf k} ^2 + m_D^2\right)^2} \,.
\label{HTLD00}
\eea

\begin{figure}[ht]
\makebox[0truecm]{\phantom b}
\put(10,40){\it (a)}
\put(10,0){\epsfxsize=6truecm \epsfbox{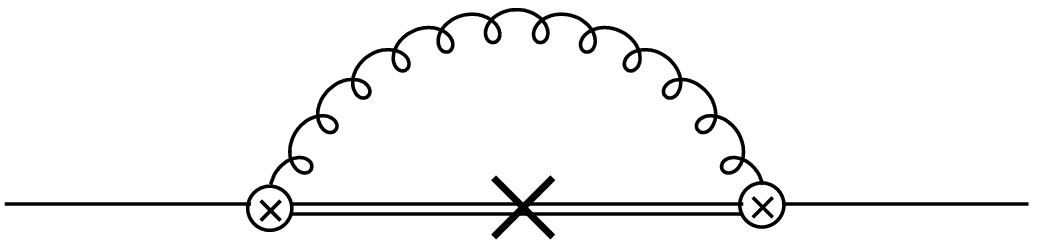}}
\put(93,-10){$V$}
\put(240,40){\it (b)}
\put(240,0){\epsfxsize=6truecm \epsfbox{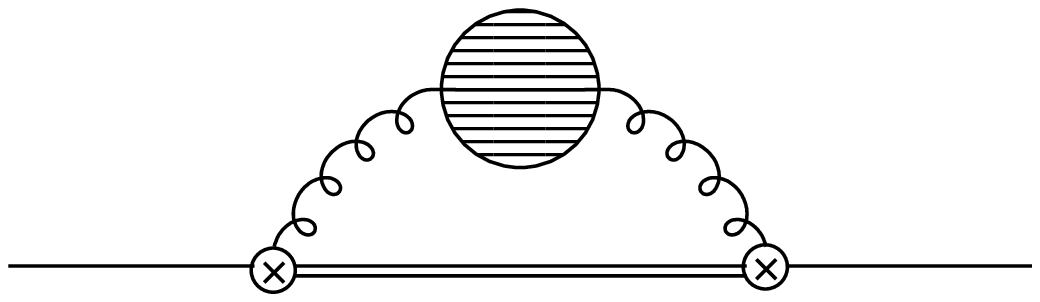}}
\caption{Diagrams contributing to the real part of the potential.
The cross in diagram {\it (a)} means that we consider only one octet potential insertion
into the free octet propagator; the shaded circle in diagram {\it (b)} stands for the gluon self-energy diagrams.}
\label{figrTVre}
\end{figure}

Also the potential in pNRQCD$_{\rm HTL}$ gets an additional thermal correction
$\delta V$ to the Coulomb potential of pNRQCD.
The leading contribution to the real part of the color-singlet potential comes from the diagrams 
shown in Fig. \ref{figrTVre} and reads
\bea
{\rm Re}~ \delta V_s(r) &=& \frac{\pi}{9} \, N_c C_F \, \als^2 \, r \, T^2 
- \frac{3}{2} \zeta(3)\,  C_F \, \frac{\als}{\pi} \, r^2 \, T \,m_D^2
+ \frac{2}{3} \zeta(3)\, N_c C_F \, \als^2 \, r^2 \, T^3 \,.
\eea
The first term stems from diagram {\it (a)} and is of order $ g^2r^2T^3\times V/T$, 
the other ones stem from diagram {\it (b)} and are of order $g^2r^2T^3\times (m_D/T)^2$. 

\begin{figure}[ht]
\makebox[0truecm]{\phantom b}
\put(10,40){\it (a)}
\put(10,0){\epsfxsize=6truecm \epsfbox{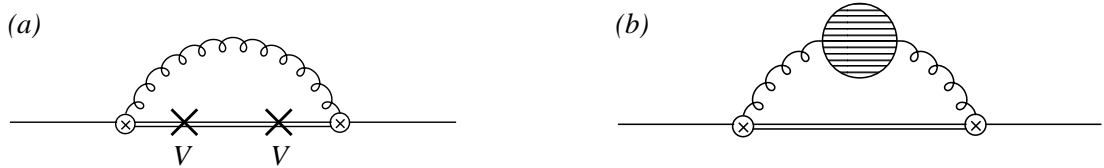}}
\put(73,-10){$V$}
\put(110,-10){$V$}
\put(240,40){\it (b)}
\put(240,0){\epsfxsize=6.5truecm \epsfbox{usmD.eps}}
\caption{Diagrams contributing to the imaginary part of the potential.
The two crosses in diagram {\it (a)} mean that we consider two octet potential insertions
into the free octet propagator.
}
\label{figrTVim}
\end{figure}

The leading contribution to the imaginary part of the color-singlet potential comes from the diagrams 
shown in Fig. \ref{figrTVim} and reads
\bea
{\rm Im}~\delta V_s(r)  &=&  - \frac{N_c^2 C_F}{6} \, \als^3\, T\,  
+ \frac{C_F}{6} \als \, r^2 \, T \,m_D^2\, \left( 
\frac{1}{\epsilon} + \gamma_E + \ln\pi 
- \ln\frac{T^2}{\mu^2} + \frac{2}{3} - 4 \ln 2 - 2 \frac{\zeta^\prime(2)}{\zeta(2)} \right)
\nn\\
& &
+ \frac{4\pi}{9} \ln 2 \; N_c C_F \,  \als^2\, r^2 \, T^3\,.
\label{VimT}
\eea
The first term stems from diagram {\it (a)} and is of order $ g^2r^2T^3\times (V/T)^2$, 
the other ones stem from diagram {\it (b)} and are of order $g^2r^2T^3\times (m_D/T)^2$. 
The imaginary part of the diagram {\it(a)} may be traced back to the singlet to octet 
break-up phenomenon introduced above while the imaginary part of the diagram {\it (b)} is due to the imaginary 
part of the gluon self energy. This may be interpreted as due to the scattering of 
soft space-like gluons emitted by the heavy quarks with hard particles (gluons and light 
quarks) in the medium. In plasma physics, this phenomenon is known as Landau damping 
\cite{Laine:2006ns,Beraudo:2007ky}. 

Divergences appear in the imaginary part of the potential at order 
$\displaystyle g^2r^2T^3\times (m_D/T)^2$, which have been regularized in 
dimensional regularization ($\epsilon = (4-d)/2$, where $d$ is the number of dimensions).
They cancel in physical observables against loop corrections from lower energy scales.
In order to illustrate the cancellation mechanism, 
let's consider the case $1/r \gg T \gg m_D \gg V$. 
Under this condition also the scale $m_D$ contributes to the potential.
Integrating out $m_D$ from  pNRQCD$_{\rm HTL}$ 
leads to an extra contribution $\delta V_s$ to the potential coming from
the diagram shown in Fig. \ref{figusT} when the momentum flowing in the loop is of 
order $m_D$ and consequently the gluon propagator is taken to be the HTL resummed 
gluon propagator as in Eq. (\ref{HTLD00}). This extra contribution reads
\bea
{\rm Re}~\delta V_s(r)  &\sim&  g^2r^2T^3\times \left(\frac{m_D}{T}\right)^3,
\label{VrealmD}
\\
{\rm Im}~\delta V_s(r)  &=&  
- \frac{C_F}{6} \, \als \, r^2 \, T \, m_D^2
\left(\frac{1}{\epsilon}-\gamma_E + \ln \pi + \ln \frac{\mu^2}{m_D^2} + \frac{5}{3} \right).
\label{VimmD}
\eea
The divergence in the imaginary part exactly cancels the one in (\ref{VimT}).

Summing up the real and imaginary parts of the potential corrections obtained 
from the scales $T$ and $m_D$, we end up with the thermal correction to the
static energy $\delta E$ and the thermal decay width $\Gamma$ respectively:
\bea
\delta E &=& 
 \frac{\pi}{9} \, N_c C_F \, \als^2 \, r \, T^2\, 
- \frac{3}{2} \zeta(3)\,  C_F \, \frac{\als}{\pi} \, r^2 \, T \,m_D^2
+ \frac{2}{3} \zeta(3)\, N_c C_F \, \als^2 \, r^2 \, T^3 ,
\\
\Gamma &=& -2\, {\rm Im}~\delta V_s = \frac{N_c^2 C_F}{3} \, \als^3\, T
\nn\\
& & - \frac{C_F}{3} \als \, r^2 \, T \,m_D^2\, \left( 2 \gamma_E 
- \ln\frac{T^2}{m_D^2} -1 - 4 \ln 2 - 2 \frac{\zeta^\prime(2)}{\zeta(2)} \right)
- \frac{8\pi}{9} \ln 2 \; N_c C_F \,  \als^2\, r^2 \, T^3.
\eea
The (leading) non-thermal part of the static energy is the Coulomb potential $-C_F \als/r$. 
The thermal width has two origins. The first term comes from the thermal break up 
of a quark-antiquark color singlet state into a color octet state. The other terms come from imaginary 
contributions to the gluon self energy that may be traced back to the Landau-damping phenomenon.
The first one is specific of QCD, the second one would also show up in QED.
Having assumed $m_D \gg V$, the term due to the singlet to octet break up 
is parametrically suppressed by $(V/m_D)^2$ with respect to the imaginary gluon 
self-energy contributions.  The $\ln {T^2}/{m_D^2}$ term is a remnant of the 
cancellation occurred between an infrared divergence at the scale $T$ and an
ultraviolet divergence at the scale $m_D$.

\section{Static quark antiquark at $T \gg 1/r \gg m_D$}
\begin{figure}[ht]
\makebox[6truecm]{\phantom b}
\put(0,0){\epsfxsize=3truecm \epsfbox{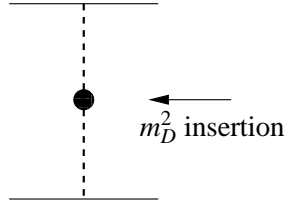}}
\put(50,25){$m_D^2$ insertion}
\caption{Leading real thermal correction to the potential. The black dot stands 
for the insertion of the real part of the HTL gluon self energy.}
\label{figpothtl}
\end{figure}

In the situation $T \gg 1/r \gg m_D$, integrating out $T$ from static QCD leads to static 
NRQCD$_{\rm HTL}$, which, at one loop, is static NRQCD with the Yang--Mills 
Lagrangian supplement by the HTL Lagrangian. 
Subsequently, integrating out $1/r$ leads to a specific version of 
pNRQCD$_{\rm HTL}$ where the Coulomb potential gets corrections from 
HTL insertions. The leading real correction comes from the diagram shown in 
Fig. \ref{figpothtl}, which gives 
\bea
{\rm Re}~ \delta V_s(r) &=&  - \frac{C_F}{2}\,\als\,r\, m_D^2 \,.
\eea
This is a correction proportional to $\als/r \times (r m_D)^2$. 

\begin{figure}[ht]
\makebox[1truecm]{\phantom b}
\put(0,40){\it (a)}
\put(20,0){\epsfxsize=4truecm \epsfbox{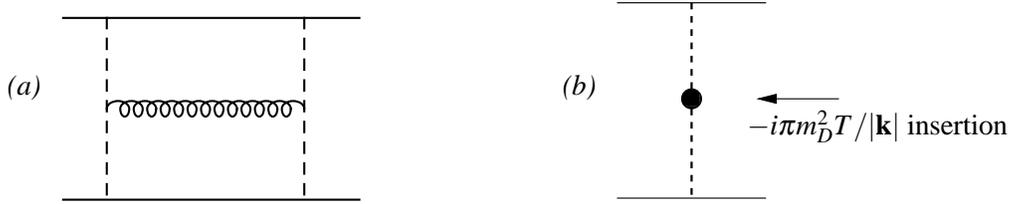}}
\put(210,40){\it (b)}
\put(230,0){\epsfxsize=3truecm \epsfbox{HTL-long-mD.eps}}
\put(280,25){$-i \pi m_D^2 T/|{\bf k}|$ insertion}
\caption{Leading contributions to the thermal decay width. Here, the black dot stands 
for the insertion of the imaginary part of the HTL gluon self energy.}
\label{figpothtlim}
\end{figure}

The leading correction to the imaginary part of the potential comes from the diagrams shown in 
Fig. \ref{figpothtlim}, which give
\bea
{\rm Im}~ \delta V_s(r) &=& 
- \frac{N_c^2 C_F}{6} \, \als^3\, T\,  
+ \frac{C_F}{6} \, \als \, r^2 \, T \, m_D^2
\left(\frac{1}{\epsilon}+\gamma_E + \ln \pi + \ln (r\,\mu)^2 -1 \right)\,.
\eea
The first term comes from diagram {\it (a)} in Fig. \ref{figpothtlim}.
It is proportional to $\als/r \times (r V)^2 \times (Tr)$ and its origin may be traced 
back to the singlet to octet break-up phenomenon. The other terms come from diagram
{\it (b)} in Fig. \ref{figpothtlim}. They are proportional to 
$\als/r \times (r m_D)^2 \times (Tr)$ and their origin may be traced back to the 
Landau-damping phenomenon.

Divergences appear in the imaginary part of the potential at order 
$\displaystyle  \als/r \times \left(r m_D \right)^2 \times (Tr)$. They cancel in physical 
observables against loop corrections from lower energy scales.
In order to illustrate the cancellation mechanism, 
let's consider the case $T \gg 1/r \gg m_D \gg V$, which is 
similar to the one discussed in the previous section.
Integrating out $m_D$ from  pNRQCD$_{\rm HTL}$ 
leads to an extra contribution $\delta V_s$ to the potential coming from
the diagram shown in Fig. \ref{figusT} when the momentum flowing in the loop is of 
order $m_D$ and consequently the gluon propagator is taken to be the HTL resummed 
gluon propagator as in Eq. (\ref{HTLD00}). These extra contributions
are the same as in Eqs. (\ref{VrealmD}) and (\ref{VimmD}).

Summing up the real and imaginary parts of the potential corrections obtained 
from the scales $1/r$ and $m_D$, we end up with the thermal correction to the
static energy $\delta E$ and the thermal decay width $\Gamma$ respectively:
\bea
\delta E &=&  - \frac{C_F}{2}\,\als\,r\, m_D^2\,,
\\
\Gamma &=& \frac{N_c^2 C_F}{3} \, \als^3\, T
+ \frac{C_F}{3} \, \als \, r^2 \, T \, m_D^2 \left(-2\gamma_E  - \ln (r m_D)^2 + \frac{8}{3} \right)\,.
\eea
The (leading) non-thermal part of the static energy is the Coulomb potential $-C_F \als/r$.
Again the thermal width has two origins. The first term comes from the thermal break up 
of a quark-antiquark color singlet state into a color octet state. The other terms come from imaginary 
contributions to the gluon self energy that may be traced back to the Landau-damping phenomenon.
Having assumed $m_D \gg V$, the term due to the singlet to octet break up 
is parametrically suppressed by $(V/m_D)^2$ with respect to the imaginary gluon 
self-energy contributions. The $\ln (r m_D)^2$ term is a remnant of the 
cancellation occurred between an infrared divergence at the scale $1/r$ and an
ultraviolet divergence at the scale $m_D$.

It is in the situation $T \gg 1/r \gg m_D \gg V$ that quarkonium in the medium
melts, if we assume that the melting condition is $E_{\rm binding} \sim \Gamma$, 
where $E_{\rm binding}$ is the quarkonium binding energy. Using the above results, the condition gives  
$g^2/r \sim g^2 T m_D^2 r^2 \, \ln 1/(m_Dr)$. For $1/r \sim m\, g^2$ and $m_D
\sim g\, T$, this leads to the melting temperature $T_{\rm melting} \sim m\, g^{4/3}\,(\ln 1/g)^{-1/3}$, 
where, assuming $g < 0.5$, we have neglected  $\ln \ln 1/g$ with respect to $\ln 1/g$
\cite{Escobedo:2008sy,Laine:2008cf}.

\section{Static quark antiquark at $T \gg 1/r \sim m_D$}
In the situation $T \gg 1/r \sim m_D$, integrating out $T$ from static QCD leads to static NRQCD$_{\rm HTL}$. 
Subsequently, both the scales $1/r$ and $m_D$ have to be integrated out at the
same time; this implies using HTL resummed gluon propagators in the matching
procedure that leads to a new specific version of pNRQCD$_{\rm HTL}$.

\begin{figure}[ht]
\makebox[0truecm]{\phantom b}
\put(20,40){\it (a)}
\put(30,0){\epsfxsize=4truecm \epsfbox{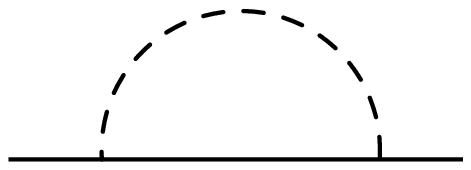}}
\put(230,40){\it (b)}
\put(250,0){\epsfxsize=3truecm \epsfbox{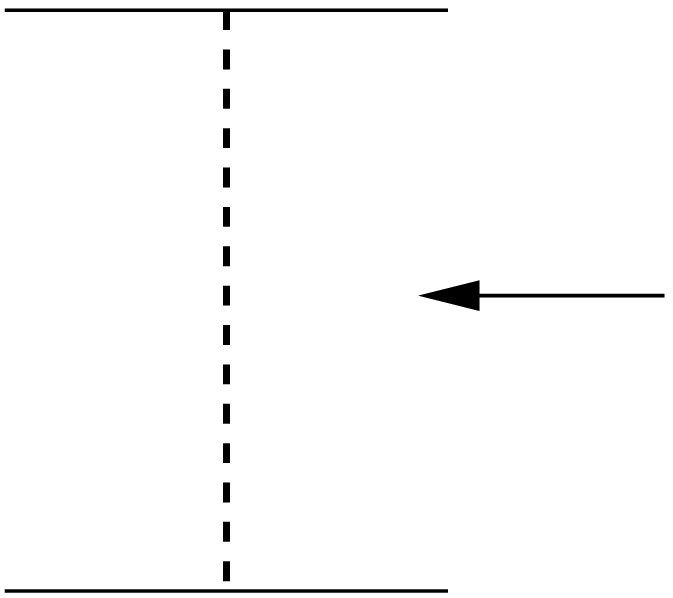}}
\put(300,25){HTL propagator}
\caption{Diagram {\it (a)} shows the leading mass self energy contribution and 
diagram {\it (b)} the leading potential contribution to the static energy.
Dashed lines stand for longitudinal HTL resummed gluon propagators.}
\label{figHTLL}
\end{figure}

The real part of the static energy is provided at leading order by the two 
diagrams shown in Fig. \ref{figHTLL}:
\bea
E = {\rm Re}~ [2\delta m + \delta V_s(r)] &=&  
 -C_F\,\als\,m_D   -C_F\,\frac{\als}{r}\,e^{-m_Dr}\,,
\eea
which is of order $\als m_D$. The result is in agreement with 
early results on $\delta m$ and $\delta V_s$ \cite{Gava:1981qd,Nadkarni:1986as}.

\begin{figure}[ht]
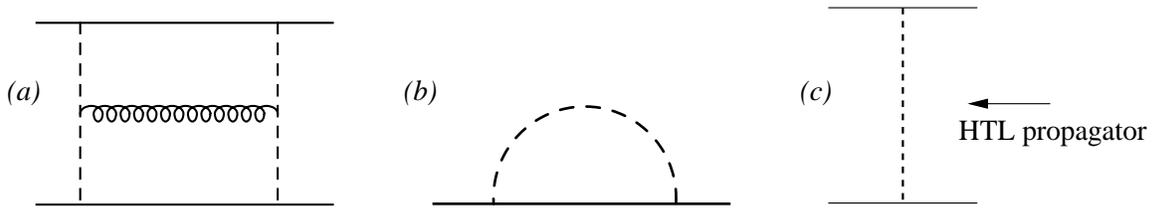

\makebox[0truecm]{\phantom b}
\put(0,40){\it (a)}
\put(10,0){\epsfxsize=4truecm \epsfbox{uswilson0.eps}}
\put(150,40){\it (b)}
\put(160,0){\epsfxsize=4truecm \epsfbox{HTL-self.eps}}
\put(300,40){\it (c)}
\put(310,0){\epsfxsize=3truecm \epsfbox{HTL-long.eps}}
\put(360,25){HTL propagator}
\caption{Diagram {\it (a)} is the leading diagram contributing to the 
singlet to octet break-up mechanism, diagram {\it (b)} contributes to the 
heavy quark damping rate and diagram {\it (c)} encodes the Landau-damping 
phenomenon. Dashed lines stand for longitudinal HTL resummed gluon propagators.}
\label{figHTLLim}
\end{figure}

The thermal decay width is  provided at leading order by the three 
diagrams shown in Fig. \ref{figHTLLim}:
\bea
\Gamma &=&  \frac{N_c^2 C_F}{3} \, \als^3\, T
+ 2\,C_F\,\als\,T \left[
1 - \frac{2}{rm_D}\int_0^\infty dx \,\frac{\sin(m_Dr\,x)}{(x^2+1)^2} \right]\,.
\eea
The first term is due to the singlet to octet break-up mechanism and is 
of order $ \als m_D \times (V r)^2 \times T/m_D$
the other ones, which were first derived in \cite{Laine:2006ns}, 
are of order $\als m_D \times T/m_D \gg \als m_D$, i.e. larger than the real
part of the energy (we recall that the binding energy is already of the
same order as the decay width at the lower temperatures discussed in the previous section).
The imaginary part of $\delta m$ is minus twice the damping rate 
of an infinitely heavy fermion \cite{Pisarski:1993rf}.

\section{Conclusions} 
In a framework that makes close contact with modern effective field theories 
of non-relativi\-stic bound states at zero temperature, we have discussed the 
real-time evolution of a static quark-antiquark pair in a medium of gluons and light 
quarks at finite temperature under the special assumption of weak coupling
both for the non-relativistic and the thermal dynamics.
For temperatures $T$ ranging from values smaller to larger 
than the inverse distance of the quark and the antiquark we have derived the
potential, the energy and the thermal decay width.  

The derived potential, $V_s$, is neither the color-singlet quark-antiquark free energy 
nor the internal energy (whose practical definition, at variance with the 
$T=0$ case, is plagued by many difficulties; 
for a recent critical discussion we refer to \cite{Philipsen:2008qx}). 
It has an imaginary part and may contain divergences
that eventually cancel in physical observables.

The derived potential describes the real-time evolution of a quarkonium state
in a thermal medium. At leading order, the evolution is governed 
by a Schr\"o\-din\-ger equation. In an EFT framework,
the potential follows naturally from integrating out all  contributions coming from modes
with energy and momentum larger than the binding energy.
For $T < V$ the potential is simply the Coulomb potential. Thermal corrections 
affect the energy and induce a thermal width to the quarkonium state; these 
may be relevant to describe the in medium modifications of quarkonium at low temperatures.
For $T >V$ the potential gets thermal contributions, which are both real and imaginary.

Two mechanisms contribute to the thermal decay width: the imaginary part of the gluon self energy 
induced by the Landau-damping phenomenon, and the quark-antiquark color singlet to color 
octet thermal break up. Parametrically, the first mechanism dominates for temperatures 
such that the Debye mass $m_D$ is larger than the binding energy, while the latter 
dominates for temperatures such that $m_D$  is smaller than the binding energy.
Finally, it has been argued that quarkonium dissociation may be a consequence of the 
appearance of a thermal decay width rather than being due to the color screening of
the real part of the potential; this follows from the observation that the
thermal decay width becomes as large as the binding energy at a temperature 
at which color screening may not yet have set in.

\acknowledgments
I thank Nora Brambilla, Jacopo Ghiglieri and P\'eter Petreczky for collaboration on the work presented here.

\end{document}